\documentclass{article}[12pt]
\usepackage{epsf}
\usepackage[latin1]{inputenc}
\usepackage{amsmath,amsfonts,amssymb}

\def\beq{\begin{equation}}
\def\eeq{\end{equation}}
\def\bea{\begin{eqnarray}}
\def\eea{\end{eqnarray}}

\def\neq{\not=}

\date{}

\begin{document}

\begin{titlepage}
\begin{center}
{\large\bf
Electron-Quasihole Duality and Second Order Differential Equation for Read-Rezayi and Jacks Wavefunctions.
}\\[.3in] 

{\bf Benoit Estienne$^{1}$, B.~Andrei Bernevig$^{2}$ and  Raoul\ Santachiara$^{3}$}\\
	$^1$ {\it Institute for Theoretical Physics, Universiteit van Amsterdam \\
Valckenierstraat 65, 1018 XE Amsterdam, The Netherlands\\
            e-mail: {\tt b.d.a.estienne@uva.nl}}\\
         
          $^2$  Department of Physics, Princeton University \\ Princeton, NJ 08544, USA \\ e-mail: {\tt bernevig@princeton.edu}\\

	$^3$  {\it LPTMS,CNRS,UMR 8626,
             Universit\'e Paris-Sud,\\ 
             B\^atiment 100,
             91405 Orsay, France. \\
    e-mail: {\tt raoul.santachiara@lptms.u-psud.fr}. }\\
\end{center}
\centerline{(Dated: \today)}
\vskip .2in
\centerline{\bf ABSTRACT}
\begin{quotation}
We consider  the quasihole wavefunctions of the non-abelian Read-Rezayi quantum Hall states which are given by the conformal blocks of  the minimal model $\textrm{WA}_{k-1}(k+1,k+2)$ of the $\textrm{WA}_{k-1}$ algebra. By studying the degenerate representations of this conformal field theories, we derive a second order differential equation satisfied by  a general many-quasihole wavefunction. We find a duality between the differential equations fixing the electron and quasihole wavefunctions. They both satisfy the Laplace-Beltrami equation.  We use this equation to obtain an analytic expression for the generic wavefunction with one excess flux.
These results also apply to the more general models  $\textrm{WA}_{k-1}(k+1,k+r)$ corresponding to the recently introduced Jack states. 
\end{quotation}
\end{titlepage}

\vskip 0.5cm
\noindent
{PACS numbers: 75.50.Lk, 05.50.+q, 64.60.Fr}

\section{Introduction}

Since the success of the Laughlin states \cite{Laughlin}, the use of trial wavefunctions in the fractional quantum Hall (FQH) effect has provided deep insights into the physics of these systems. Recent new interest has focused on non-Abelian Fractional Quantum Hall states of matter.  
These are gapped phases of matter  in which the adiabatic transport of one fundamental excitation around another implies a unitary transformation within a subspace of degenerate wavefunctions which differ from each other only globally \cite{RM_review}. Systems exhibiting non-Abelian statistics can store topologically protected qubits and are therefore interesting for topological quantum computation \cite{Sarma}.

Much of the comprehension of non-Abelian quantum Hall states  relies on the conformal field theory (CFT) \cite{diFrancesco,Dotsi_cours} approach . In this approach, the analytic part of trial wavefunctions is given by  the conformal blocks of a given CFT \cite{MooreRead,ReadRezayi}.  The basic idea  is that the effects of an  adiabatic exchange  of excitations coincides with the monodromy properties of conformal blocks \cite{RM_review,Read_holonomy}.  In this case, the universal properties characterizing a topological phase, such as the quantum numbers of the ground state or the statistics of the excitations, are directly related to the analytic properties of the conformal blocks in the corresponding CFT.

The Read-Rezayi (RR) states play a paradigmatic role in the physics of non-Abelian FQH  states. The RR states are built from the conformal blocks of the $\mathbb{Z}_k$ parafermionic theory \cite{FZ} and describe bosons (fermions) at Landau level filling fractions $\nu=k/(M k +2)$, where $M$ is an even (odd) integer. For exemple, the $k=1$ RR states correspond to the simplest Laughlin states associated to the (abelian) $c=1$ CFTs. The $k=2, M=1$ RR state corresponds to the Moore-Read state \cite{MooreRead} which describes  a (p-wave) pairing of electrons occuring  at  the first excited  Landau level for filling fraction $\nu=5/2$. The non-Abelian nature of the Moore-Read state  is  well understood in terms of the $\mathbb{Z}_2$ parafermionic theory, which coincides with the Ising CFT \cite{diFrancesco,Dotsi_cours}. 
The general $k$ RR states \cite{ReadRezayi} have been introduced as a natural generalization of the pairing $k=2$ structure  of the Moore-Read states. The $k$ RR states describe incompressible fluids made by clusters of $k$ particles and can be fully characterized  by  specifying the manner in which these states vanish as $k+1$ particles come to the same point.  This  so-called $k-$clustering properties are inherited  from the $\mathbb{Z}_k$ symmetry of the parafermionic theory and   makes the $k$ RR  states to be the exact zero-energy states of a $k+1$ body interaction Hamiltonians\cite{ReadRezayi,SRC_projham}.  It is now established that the RR sequence could be  physically relevant:  the $k=3,4, M=1$ RR state, for instance, may  capture the physics of the FQH plateau observed at $\nu=12/5$ and $\nu=13/5$\cite{Read_Rezayi2} while bosonic RR states may also be realizable in rotating Bose gases \cite{Cooper_bos}. Moreover, for topological quantum computation purposes, the $k>2$ RR states are particularly interesting as their non-abelian braiding is sufficiently rich  to carry out universal quantum computation \cite{Bonesteel}.

Since their introduction, many properties of the RR states, such as the degeneracy of the quasihole excitations and the associated braiding properties, have been studied in detail. Generally, these properties can be derived  from the parafermionic fusion rules which in turn can be directly read out, together with the values of the conformal dimension of the fields, from  the  representation theory of the $\mathbb{Z}_k$ chiral algebra. On the other hand, the computation of multi-point correlation functions or, in other words,  of the explicit expressions of the wavefunctions, is much more difficult.  Clearly, the knowledge of the exact expression of the wavefunctions provides information which goes well beyond the quasihole braiding properties, as been discussed in \cite{Nayak_Wilczek,Ardonne_Schoutens}. For the RR states   \cite{ReadRezayi}, the form of the ground state wavefunction  is known explicitly for any number of particles as being a Jack polynomial. Concerning the excited wavefunctions, an explicit expression is not known. An implicit construction of the linearly independent quasihole polynomials is given in  \cite{Read_2} but the explicit expression for the state with one flux added in the general $\mathbb{Z}_k$ theory is still missing.  The simplest  excited wavefunctions, i.e. with the smallest number of quasiholes  exhibiting non-Abelian statistics have been evaluated in \cite{Nayak_Wilczek} for the $k=2$ Moore-Read case and in \cite{Ardonne_Schoutens} for general $k$. 
Still, the explicit expressions for general quasihole wavefunctions and for  general $k$ is not known.

It is the purpose of this paper  to find  the differential operator which annihilates a general quasihole wavefunction of the RR series and use it to obtain a general expression for the excited one flux-added excitations of al the $\mathbb{Z}_k$ RR states. During the process, we find that the Read-Rezayi pinned quasihole states (and in fact all the the Jack polynomial states) satisfy a previously unknown electron-quasihole duality. We find that the differential operator that diagonalizes the pinned quasihole wavefunction  is second order in \emph{both} electron and quasihole coordinates. Moreover, the form of the operator in electron and quasihole coordinates is of Laplace Beltrami form but with  dual coupling strength. We use this to obtain the expression for the  one-flux-added  quasihole wavefunction for all the $\mathbb{Z}_k$ Read-Rezayi states. Due to the dual nature of the electron and quasihole states, the wavefunction for pinned quasiholes decomposes into a sum over products of Jack polynomials in electron coordinates and dual Jack polynomials in quasihole coordinates.

A possible way to derive differential equations for RR states  would be  to use the  $SU(2)_k/U(1)$ coset realization of the $\mathbb{Z}_k$ parafermionic theory.  Indeed, the correlation functions of the $\mathbb{Z}_k$ parafermionic theory can be directly related to the ones of the $SU(2)_k$  primaries  of a Wess-Zumino-Witten model \cite{Witten} which satisfy a system of linear differential equations, the so called Knizhnik-Zamolodchikov equations \cite{kz}. However, this method is increasingly impractical as the size of the system of differential equations grows with the number of quasiholes and particles.  While in small number of  cases (such as the $\mathbb{Z}_3$ and $\mathbb{Z}_2$ parafermions) the differential equation satisfied by the quasiholes can be obtained through identifying the theory with a minimal model, it is generally of  high-order in derivatives (third order for the $\mathbb{Z}_3$ RR state) . The differential equation we obtain contains only second order derivatives for any $k$. The lowering of derivative order is important - the fact that the Laughlin or Halperin wavefunctions are annihilated by first-order differential operators is crucial to the mathematical proof of the quasihole braiding.

 Our  starting point will be instead to consider the description of the $\mathbb{Z}_k$ parafermionic theory  in terms of a  family of CFTs  with $W$ extended symmetry, the $\textrm{WA}_{k-1}$ CFTs \cite{Wtheory,Schoutens}. The importance of the $W$ symmetry for understanding the RR sequence was pointed  out and used in \cite{Cappelli}.   We will show in this paper that, by exploiting the $W$ symmetry, we can derive a   second-order differential equation satisfied  by  a general $k$ RR quasihole wavefunction. Moreover, our approach can be straightforwardly applied to a more general family of trial wavefunctions which can be written in term of single Jack polynomial  \cite{McDonald} and therefore named Jacks wavefunctions\cite{BernevigHaldane1,BernevigHaldane2}.  
 
 The series of minimal models $\textrm{WA}_{k-1}(p,q)$ associated to the  $\textrm{WA}_{k-1}$ algebras are indexed by two coprime integers $p$ and $q$ \cite{Wtheory,Schoutens}. The $\mathbb{Z}_k$ parafermionic theory  coincides with the theory $\textrm{WA}_{k-1}(k+1,k+2)$. The Jack wavefunctions represent a  possible generalization   of the Read-Rezayi states based on the $\textrm{WA}_{k-1}(k+1,k+r)$ theory, with $r>2$ \cite{BernevigHaldane1,BernevigHaldane2,Ardonne,Mathieu,TJ}.  The states constructed from the $\textrm{WA}_{k-1}(k+1,k+r)$ conformal block  have been shown to satisfy the so called $(k,r)$ clustering conditions \cite{ BernevigHaldane1,BernevigHaldane2,FJMM,FJMM2}: they vanish with power $r$  when  the $k+1^{\textrm{th}}$ particle approches a cluster of $k$ particles. Here, it is important to stress that the $\textrm{WA}_{k-1}(k+1,k+2)$ are non-unitary for $r>2$. Although  there is increasingly strong evidence that only rational unitary CFTs can describe a gapped phase of matter\cite{Readnu1,Readnu2,Thierry}, these non-unitary states may still describe interesting critical points between gapped phases and therefore be worth studying \cite{BernevigHaldane1,BernevigHaldane2,Gaffnian,BernevigHaldane3,BernevigW}. Unitary CFTs generating $(k,r)$ clustering polynomials have been discussed in \cite{ERS} and their physical relevance further discussed in \cite{S3}.

In \cite{RB1}, two of us  considered the ground state wavefunctions for general  $\textrm{WA}_{k-1}(k+1,k+r)$
states. In this respect we studied the multi-points correlation functions of certain $W$ primary fields which are identified as particle  operators (defined below). By exploiting the representations of the $\textrm{WA}_{k-1}$ symmetry, and in particular the degeneracy properties of these particle fields, we showed that their $N-$point  correlation functions  satisfy a second order differential equation. Moreover we showed that this equation can to be transformed into a Calogero Hamiltonian with negative rational coupling. This completed  the proof of the fact that the ground states $\textrm{WA}_{k-1}(k+1,k+r)$ theories can be written as a single Jack polynomial  \cite{BernevigHaldane1,BernevigHaldane2}. 
In this paper we show that the approach followed in \cite{RB1} can be extended to the study of general quasihole wavefunctions for the Read-Rezayi states and, more generally, to the quasihole wavefunctions of the $\textrm{WA}_{k-1}(k+1,k+r)$ theories. Moreover, we show how to apply the second-order differential equation to  rigorously generalize an ansatz found in earlier works \cite{BernevigHaldane2} for the computation of certain quasihole wavefunctions. In the process we find a previously unknown electron-quasihole duality.

In section 2 we define the FQH trial wavefunctions which are the object of our study and we make explicit the FQH-CFT connection. In section 3, we characterize completely the electron and quasihole operators by specifying their $\textrm{WA}_{k-1}$ quantum numbers. In section 4 we derive a partial differential equation satisfied by the quasihole wavefunctions. In section 5 we present the main results of the paper: we derive the PDE for generic quasihole wavefunctions describing the non-abelian Jack quantum Hall states. This PDE turns out to be surprisingly simple, and exhibits an interesting duality between electrons and quasiholes. In section 6 we solve the PDE for the case of one flux-added wavefunctions and obtain them as a nice expansion of products of Jack polynomians in electron and quasihole coordinates. Section 7 gives our conclusions, an Appendix 8 gives a wavefunction example.

\section{General form of Read-Rezayi and Jacks wavefunctions}
In this section we briefly define the FQH trial wavefunctions which are the object of our study and we make explicit the FQH-CFT connection.

It is convenient to consider a system of $N$ particles on a sphere of radius $R$ with a uniform radial magnetic field with total flux $N_{\phi}$ \cite{Haldane_sphere}.  The position of $i^{\textrm{th}}$ particle on the sphere can be represented as a complex variable $z_i$ which is its stereographic projection. Each particle in the lowest Landau level has orbital angular momentum $N_{\phi}/2$ and the single-particle basis states have the form $z^m \mu(z,\bar{z})$, $m=0, \dots N_{\Phi}$. The $L_z $ momentum quantum number is $m-\frac{N_{\Phi}}{2}$ and $\mu(z,\bar{z})$ is the measure on the sphere, $\mu(z,\bar{z})=1/(1+(|z|/2R)^2)^{1+N_{\phi}/2}$. Therefore a many-body  wavefunction $\tilde{\Psi}$\footnote{we use $\tilde{\Psi}$ to distinguish  the many-particles wavefunction from the parafermionic field $\Psi$ defined later} describing $N$ particle states in the lowest Landau levels takes  the form:
\begin{equation}
\tilde{\Psi}(z_1,\bar{z}_1,\cdots,z_N, \bar{z}_N)=P_{N}(z_1,\cdots,z_n)\prod_{i=1}^{N}\mu(z_i,\bar{z}_i),
\end{equation}
where $P_{N}(\{z_i\})$ is a polynomial in the $N$ variables $z_i$, (anti)symmetric for bosons (fermions).  In the following we drop the measure factors and focus on the analytic part of the many-body wavefunction.

\subsection{Ground state wavefunction}

A quantum Hall ground state wavefunction has to be rotationally invariant. This means that the corresponding polynomial $P_{N}(\{z_i\})$ is  translationally  invariant and homogeneous with a total degree $\frac{1}{2}NN_{\Phi}$, with $N_{\phi}$ being the highest power in each variable $z_i$. 
On the sphere, $N_\phi$ and $N$ are related by the linear identity
\begin{equation}
N_\phi=\nu^{-1}N-\delta
\end{equation}  where $\nu$ is the filling fraction and $\delta$ the so-called shift. 
 Let $P_{N}^{(k,r)}(z_1,\cdots,z_n)$ denote a symmetric polynomial satisfying the $(k,r)$ clustering properties defined by:
\begin{equation}
P_{N}^{(k,r)}(\underbrace{z_1=\dots=z_{k}}_{\text = Z},z_{k+1},z_{k+2},\cdots, z_{N})=\prod_{i=k+1}^{N}(Z-z_i)^r P_{N-k}^{(k,r)}(z_{k+1},z_{k+2},\cdots, z_{N})\label{rpower}
\end{equation}
Because of the above properties, the state $P_{N}^{(k,r)}(\{z_i\})$ is a zero energy eigenstates of the projection operator $\mathcal{P}^{r}_{k}$ \cite{SRC_projham} which  annihilates states where any cluster of $k+1$ particles has relative  angular momentum less than $r$. In other words no more than $k$ particles can occupy  $r$ consecutive orbitals.  A (partial) classification of the symmetric polynomial with $(k,r)$ clustering conditions has been discussed in \cite{Wen1,Wen2,Wen3}.

 \subsubsection{$(k,2)$ clustering states: the Read-Rezayi sequence}
 
 The RR sequence describes particles at filling fraction
 \begin{equation}
 \nu=k/(2+k M) \quad \delta=2+M
 \label{fillingRR}
 \end{equation} 
Hereafter, without any loss of generality,  we focus on the  bosonic $M=0$ RR states.  The RR states with general $M$ are related to the case $M=0$ by a simple Jastrow factor.

The  $k$ RR ground state  is uniquely defined by the $(k,2)$ clustering conditions \eqref{rpower}.
The polynomial $P_{N}^{(k,2)}(\{z_i\})$ can be constructed in the following way:
\begin{equation}
P_{N}^{(k,2)}(z_1,\cdots,z_N)=\langle \Psi_1(z_1) \hdots  \Psi_1(z_{N}) \rangle \prod_{i<j} \left( z_i-z_j \right)^{2/k}  
\label{RR_paraconn}
\end{equation}
where $\langle \Psi_1(z_1) \hdots  \Psi_1(z_{N}) \rangle$ is the $N-$points conformal correlation function of the parafermionic field $\Psi_1$. The field $\Psi_1$ has been identified as the particle operator. This identification traces back to the fact that  the field $\Psi_1$  is the fundamental conserved current associated  to the extended $\mathbb{Z}_k$ symmetry.  The associated chiral algebra is formed by the set of chiral currents $\{ \Psi_{0}(z), \Psi_{1}(z), \cdots \Psi_{k-1}(z)\}$, $\Psi_{0}=I$ being the identity operator. These currents have conformal dimensions:
\begin{equation}
\Delta_{q}=\frac{q(k-q)}{k}, 
\label{dim_rr}
\end{equation} 
and satisfy the following operator product expansions (OPEs):
\begin{eqnarray}
\Psi_q(z) \Psi_{q'}(w)&=&\frac{\gamma_{q,q'}}{(z-w)^{\Delta_{q}+\Delta_{q'}-\Delta_{q+q'}}}[\Psi_{q+q'}(w)]
\label{para_algebra1}\\
\Psi_q(z) \Psi_{k-q}(w)&=&\frac{1}{(z-w)^{2\Delta_{q}}}\left( 1+(z-w)^2\frac{2\Delta_{q}}{c} T(w)+\cdots \right)
\label{para_algebra2}
\end{eqnarray}
where the sums $q+q'$ are defined modulo $k$, $[\Psi]$ indicates the operator $\Psi$ and its descendants, while the $\gamma_{q,q'}$ are the algebra coupling constants. Notice that the fields $\Psi_{1}$ and its conjugate $\Psi_{k-1}$ have the same conformal dimension and  fuse to the identity, $\Psi_{1} \times \Psi_{k-1}= 1$.
The field $\Psi_{k-1}(z)$ represents a  cluster  of $k-1$ particles at the position $z$.

The  $(k,2)$ clustering properties \eqref{rpower} are a direct consequence of the fusion rules \eqref{para_algebra1} and \eqref{para_algebra2}. Moreover the relations \eqref{RR_paraconn} and \eqref{dim_rr} imply $N_{\phi}=2(N-k)/k$, which corresponds to Eq.\eqref{fillingRR} with $M=0$.

\subsubsection{$(k,r)$ clustering states: Jacks wavefunctions}
 
The OPEs \eqref{para_algebra1}-\eqref{para_algebra2} are the general form for a candidate chiral algebra realizing the $\mathbb{Z}_k$ symmetry. 
When considering arbitrary values of the conformal dimensions $\Delta_q$, the principal difficulty is to define completely the form of the corresponding OPEs, i.e. all the singular terms, in order to obtain an associative algebra.  Other associative solutions besides the ones in \eqref{dim_rr}  have been found for the following values \cite{zomo32,DotsenkoRaoul,DotsenkoRaoul2,JacobMathieu1,JacobMathieu2} :
\begin{equation}
\Delta_{q}=\frac{r}{2}\frac{q(k-q)}{k},
\label{dim_jj}
\end{equation}
where $r$ is an integer ($r=2,3,\dots$). In the following we denote this algebra as $\mathbb{Z}_{k}^{(r)}$. 

The  $\textrm{WA}_{k-1}(k+1,k+r)$ conformal field theory is a particular realization of this algebra. It contains  a set of primaries $\Psi_q$ of dimension \eqref{dim_jj} that forms a parafermionic algebra \eqref{para_algebra1}-\eqref{para_algebra2} \cite{Wtheory}   Let $\langle \Psi_1(z_1) \hdots  \Psi_1(z_{N}) \rangle^{(r)} $ denote the correlator of $N$  parafermionic fields $\Psi_1$  of the $\textrm{WA}_{k-1}(k+1,k+r)$ theory. The OPEs \eqref{para_algebra1}-\eqref{para_algebra2} together with the conformal dimensions \eqref{dim_jj} imply that  the polynomial
\begin{equation}
P_{N}^{(k,r)}(z_1,\cdots,z_N)\equiv \langle \Psi_1(z_1) \hdots  \Psi_1(z_{N}) \rangle^{(r)} \prod_{i<j} \left( z_i-z_j \right)^{r/k} , 
\label{JJ_paraconn}
\end{equation}
satisfies the  $(k,r)$ clustering properties. Moreover, these clustering states  describe particles at filling fraction \cite{BernevigHaldane1,BernevigHaldane2}
\begin{equation}
 \nu^{(r)}=k/r \qquad \textrm{and} \qquad \delta^{(r)}=r
 \label{fillingJJ}
 \end{equation} 
Clearly, for $r=2$, one finds the quasihole wavefunctions of the RR sequence. In \cite{RB1} it was proven that the ground state wavefunctions $P_{N}^{(k,r)}$ defined in \eqref{JJ_paraconn} can be written in terms of single Jacks polynomials (see section \ref{PDE}). The aim of this paper is to extend this analysis to quasihole states.

\subsection{Quasihole wavefunctions}
\label{quasihole}

One type of excitations over the quantum Hall ground state are obtained by injecting  quasiholes, i.e. defect of charge, at fixed  positions $w_j$.  The  elementary quasihole operator is represented generally by  a field $\sigma$ which has to be identified among the primaries of the correspondent CFT theory. In order to identify the quasihole operator, it is required that particles  and quasihole be mutually local.

We consider in the following a general $\textrm{WA}_{k-1}(k+1,k+r)$ theory, recovering the RR states by setting $r=2$. The field representing the elementary quasihole operator for the RR states and more generally for $(k,r)$ clustering Jack states is a field $\sigma$ with dimension $h$. 
\begin{equation}
h=\Delta_{\sigma}=\frac{(k-1)(1+k(2-r))}{2 k(r+k)}
\label{hsigma}
\end{equation}
The conformal dimension $h$ has been extracted in  \cite{BernevigW} from the clustering properties, and the assumption that the electronic wavefunctions can be described purely in terms of $(k,r)$ admissible Jack polynomials.

The OPEs between the fields $\Psi_{1}$ and $\Psi_{k-1}$ and $\sigma$ have the form:
\begin{eqnarray*}
\Psi_{1}(z)\sigma(w)&=&\frac{1}{(z-w)^{1/k}}\phi(w)+\cdots  \\
\Psi_{k-1}(z)\sigma(w)&=&\frac{1}{(z-w)^{(k-1)/k}}\phi'(w)+\cdots 
\label{fuspsiphi}
\end{eqnarray*} 
where $\phi$ and $\phi'$ are others $\textrm{WA}_{k-1}(k+1,k+r)$ primaries. In the Hilbert space of the  $\textrm{WA}_{k-1}(k+1,k+r)$ one finds the field conjugate to $\sigma$. This field $\sigma'$ has the same conformal dimension  $\Delta_{\sigma'}=h$, see \eqref{hsigma}, and has a fusion channel with $\sigma$ into the identity:
\begin{equation}
\sigma(z) \sigma'(w)=\frac{1}{(z-w)^{2 h}} +\cdots \label{fusphiphi}
\end{equation} 
The field $\sigma'$ represents then the object which is precisely the fusion of $k-1$ quasiholes. The fusion rules of the field $\sigma'$ with the fields $\Psi_{1}$ and $\Psi_{k-1}$ have the form:
\begin{eqnarray*}
\Psi_{1}(z)\sigma'(w)&=&\frac{1}{(z-w)^{(k-1)/k}}\tilde{\phi}'(w)+\cdots  \\
\Psi_{k-1}(z)\sigma'(w)&=&\frac{1}{(z-w)^{1/k}}\tilde{\phi}(w)+\cdots 
\label{fuspsiphip}
\end{eqnarray*} 
Again the fields $\tilde{\phi}$ and $\tilde{\phi}'$ are $W$ primaries with the same dimension respectively of the fields $\phi$ and $\phi'$ appearing in \eqref{fuspsiphi}.
 The degenerate subspace of wavefunctions with $M$ quasiholes  is then built of the different conformal blocks of the correlator:
\begin{eqnarray}
P_{M,N}^{(k,r)}(w_1,w_2,\cdots, w_M|z_1,\dots,z_N)&=&\langle \sigma(w_1) \sigma(w_2)\cdots \sigma(w_M) \Psi(z_1) \cdots \Psi(z_N) \rangle^{(r)}_{(a)}\times \nonumber \\&& \times \prod_{i<j}^{N}(z_i-z_j)^{\frac{r}{k}} \prod_{i=1}^{N}\prod_{j=1}^{n} (z_i-w_j)^{\frac{1}{k}} \prod_{i<j}^{n}(w_i-w_j)^{\frac{1}{2 r k}}. \label{conformal block label}
\end{eqnarray}
In the above expression the index $r$ indicates that we are considering a correlation function of the $\textrm{WA}_{k-1}(k+1,k+r)$ theory while the index $a$ runs over the possible conformal blocks.

\section{$\textrm{WA}_{k-1}$ theories: characterization of the electron and quasihole operators}
\label{wtheory}
In this section we characterize completely the electron  and quasihole operators by specifying their $\textrm{WA}_{k-1}$ quantum numbers. 
 
\subsection{Representation theory of the $\textrm{WA}_{k-1}$ algebras: main results}
A complete construction of $W$ algebras  and their representation theories can be found  in \cite{Wtheory, Schoutens}.  A brief review of the main results about $W$ theories  can be found in \cite{RB1}.  In the following we just report the main definitions about $W$ theories and we fix our notations.

The chiral fields $W^{(s)}(z)$ ($s=3,\cdots, k-1$) classify, together with $T(z)$, all the operators  of the model in terms of primaries and descendants of the chiral algebra $\mathcal{W}_k$.  The  primary fields $\Phi_{\vec{\beta}}$ of the theory are  parametrized by the $k-1$  component vector $\vec{\beta}$. The behavior of a primary field $\Phi_{\vec{\beta}}$ under the action of the symmetry  generators $W^{(s)}(z)$ is encoded in the OPEs:
\begin{eqnarray}
  T(z)\Phi_{\vec{\beta}} (w) =   \sum_{n=-\infty}^{\infty} \frac{L_{n}\Phi_{\vec{\beta}} (w)}{(z-w)^{n+2}} & = &\frac{\Delta_{\vec{\beta}} \Phi_{\vec{\beta}}(w)}{(z-w)^2 } + \frac{\partial_{w}\Phi_{\vec{\beta}}(w) }{(z-w)} +\cdots  \\
   W^{(s)}(z)\Phi_{\vec{\beta}} (w) =  \sum_{n=-\infty}^{\infty} \frac{W_{n}^{(s)}\Phi_{\vec{\beta}} (w)}{(z-w)^{n+s}} & =& \frac{\omega^{(s)}_{\vec{\beta}}\Phi_{\vec{\beta}}(w)}{(z-w)^s } +\frac{W^{(s)}_{-1}\Phi_{\vec{\beta}}(w)}{(z-w)^{s-1}}+ \frac{W^{(s)}_{-2}\Phi_{\vec{\beta}}(w)}{(z-w)^{s-2}}  + \cdots
 \label{Wprimary} 
\end{eqnarray}   
The conformal  dimension $\Delta_{\vec{\beta}}$ and the $k-2$ quantum numbers $\omega^{(s)}_{\vec{\beta}}$ are respectively the zero mode eigenvalues of the chiral fields $T(z)$ and $W^{(s)}(z)$: 
\begin{equation}
L_{0}\Phi_{\vec{\beta}} = \Delta_{\vec{\beta}} \Phi_{\vec{\beta}} \qquad W^{(s)}_{0}\Phi_{\vec{\beta}}= \omega^{(s)}_{\vec{\beta}}\Phi_{\vec{\beta}} \qquad s=3,\cdots, k-1
\end{equation}
They characterize each  representation $\Phi_{\vec{\beta}}$. In particular the conformal dimension $\Delta_{\vec{\beta}}$ is given by:
\begin{equation}
\Delta_{\vec{\beta}}=\frac{1}{2}\vec{\beta}(\vec{\beta}-2\vec{\alpha}_{0})
\end{equation}
Notice also that, from the above definitions, $L_{-1}\Phi_{\vec{\beta}}(z)=\partial_{z} \Phi_{\vec{\beta}}(z)$.

The allowed values of the vectors $\vec{\beta}$ are defined by the
condition of complete degeneracy of the modules of $\Phi_{\vec{\beta}}(z)$ with
respect to the chiral algebra. 
 The Kac table is based on the weight lattice of the Lie algebra $A_{k-1}$ and the position of 
 the vectors $\vec{\beta}$ are found to be given by \cite{Wtheory}:
\begin{equation}
\vec{\beta}=\sum_{a=1}^{k-1}\left((1-n_{a}) \alpha_{+} + (1-n_{a}')\alpha_{-}\right)\vec{\omega}_{a} \equiv \left(n_1,n_2\cdots n_{k-1}  \mid n_1',n_2'\cdots n_{k-1}'\right)
\end{equation}
Each primary operator $\Phi_{(n_1,\cdots,n_{k-1}|n_1',\cdots,n_{k-1}')}$ is then labeled by a set of integers $n_a,n'_a$ ($a=1,\cdots, k-1$).
The principal domain of the Kac table contains the set of primary operators which form a closed fusion algebra, and is delimited as follows:
\begin{eqnarray}
n_a \geq 1 & ; & n'_a \geq 1 \\
\sum_{a} n_a  <  p' & ; & \sum_{a} n_a'  <  p 
\label{Kac}
\end{eqnarray}
Introducing $n_0 \equiv p'-\sum_{a=1}^{k-1} n_{a}$ and $n'_0 \equiv p-\sum_{a=1}^{k-1} n'_{a}$, the Kac table is simply delimited by $n_a, n'_a \geq 1$ ($a=0,\cdots, k-1$).
Moreover, positions in the Kac table are identified modulo cyclic permutations of the indices:
\begin{eqnarray}
n_a \to n_{a+1} \quad  \textrm{and} \quad n_{a}' \to n_{a+1}'   \qquad \textrm{with} \quad n_k \equiv n_0
\label{identifications}
\end{eqnarray}
This means that the fields $ \Phi_{(n_1,\cdots,n_{k-1}|n'_1,\cdots,n'_{k-1})}$  and $\Phi_{(n_2,\cdots,n_{k-1},n_0|n'_2,\cdots,n'_{k-1},n'_0)}$  represent the same primary field, and implies that they have the same quantum numbers $\omega^{(s)}$. In particular the conformal dimension is invariant: 
\begin{equation}
\Delta_{(n_1,\cdots, n_{k-1}|n_1'\cdots n'_{k-1})} = \Delta_{(n_2,\cdots,n_{k-1},n_0|n'_2,\cdots,n'_{k-1},n'_0)}
\end{equation}
The  representation corresponding to the primary field $\Phi_{(\vec{n} | \vec{n}')}$
exhibits $k$ null vectors $\chi_{a}$ ($a=0,\cdots, k-1$) at level $N_a=n_{a}n_{a}'$. This directly generalizes the well known case of the degenerate representations of the Virasoro algebra (\emph{i.e.} the $\textrm{WA}_1$ algebra) \cite{diFrancesco}. Another remark is that the primary fields $\Phi_{(\vec{n} | \vec{n}')}$ and $\Phi_{(\vec{n}' | \vec{n})}$ are dual from one another, in the sense that the corresponding modules have the same degeneracies. This will translate into the FQHE language as an electron-quasihole duality, and will play a major role in the discussion of quasihole wavefunctions.

\subsection{$\textrm{WA}_{k-1}$ quantum numbers of electron and quasihole operators.}

Of particular interest is the model $\textrm{WA}_{k-1}(k+1,k+r)$ where $p=k+1$ and $p'=k+r$. By using the fusion rules of this theory \cite{Ardonne,RB1}, one can verify that the set of operators:
\begin{equation}
\Psi_{i}= \Phi_{-\alpha_-\vec{\omega}_{i}} = \Phi_{-r\alpha_+\vec{\omega}_{k-i}}  \quad i=1,\dots,k-1, \quad \Delta_{i}=\frac{r}{2}\frac{ i(k-i)}{k}
\label{parafermionic_operators}
\end{equation}
form a parafermionic algebra \eqref{para_algebra1}-\eqref{para_algebra2}, namely $\Psi_{i}\times \Psi_{j}=\Psi_{i+j \;\mbox{mod}\;k}$. An important remark is that these fusion rules are only valid when $p=k+1$. In general the fusion rules of these fields are multi channeled, but for $p=k+1$ the identifications \eqref{identifications} gives:  
\begin{equation}
\Psi_{i} = \Phi_{\substack{(1,\dots,1 | 1,\dots ,1,2,1,\dots ,1)\\ \phantom{(1,\dots,1 | 1,\dots ,1,}\uparrow\phantom{,1,\dots ,1)} \\\phantom{(1,\dots,1 | 1,\dots ,1,}i^{\textrm{th}}\phantom{,1,\dots ,1)}}} = \Phi_{\substack{(1,\dots ,1,r+1,1,\dots ,1|1,\dots,1)\\ \phantom{(1,\dots ,1,}\uparrow\phantom{,1,\dots ,1|1,\dots,1)} \\ \phantom{(1,\dots ,1,}k-i^{\textrm{th}}\phantom{,1,\dots ,1|1,\dots,1)} }} \label{parafermionic_operators_index}
\end{equation}
and the usual fusion rules are truncated accordingly, becoming effectively single channeled.
 
Notice that the correlation function of $\Psi_{i}$ operators are symmetric under the conjugation of charge $i\to k-i$. This means that one could take $\Psi_{k-1}$ as the electron operator instead. In particular the conformal dimension is insensitive to this charge conjugation. However the eigenvalue of $W^{(3)}_0$  changes sign under  $i\to k-i$, and two conjugate fields have opposite value of the quantum number  $\omega^{(3)}$, as can be derived from the Ward identities of the two-point function. 

The eigenvalues associated to the $\Psi_{1}$ and $\Psi_{k-1}$ representations are fixed to:
\begin{equation}
(\omega^{(3)}_{\Psi_1})^2= (w^{(3)}_{\Psi_{k-1}})^2=\frac{(k-1)^2(k-2)}{18 k^3} \frac{r^2(r(k+2)+k)}{(3k+2)-rk}
\end{equation}
From the above equations, one can see that  the signs of $\omega^{(3)}_{\Psi_{1}}$  and  $\omega^{(3)}_{\Psi_{k-1}}$ are not fully determined.
There is a global ambiguity in the sign for the quantum number  $\omega^{(3)}$, as one is free to change the sign of the current $W^{(3)}(z)$ without changing the  OPEs of the $\textrm{WA}_{k-1}$ algebra. This is not true for the conformal dimension as the sign of the stress energy tensor $T(z)$ is set by the OPE:
\begin{equation}
T(z)T(0) = \frac{c/2}{z^4} + \frac{2T(0)}{z^2} + \frac{\partial T(0)}{z} + \cdots
\end{equation}
In the following we choose a sign for $\omega^{(3)}_{\Psi_{1}}$, and all the remaining eigenvalues are then fully determined. We use quite often the symbol  $\Psi$ for the electron field $\Psi_{1}$,  $\Delta$ for its conformal dimension:
\begin{equation}
\Delta=\frac{r}{2}\frac{(k-1)}{k}
\label{Dpsi}
\end{equation}
and $\omega_{\Psi}^{(3)}$ for the associated  eigenvalue  of $W^{(3)}_0$.  We set then:
\begin{equation}
\omega^{(3)}_{\Psi} = - \omega^{(3)}_{\Psi_{k-1}} =\sqrt{\frac{(k-1)^2(k-2)}{18 k^3} \frac{r^2(r(k+2)+k)}{(3k+2)-rk}} 
\label{w3psi}
\end{equation}
The quasihole operators  $\sigma$ and $\sigma'$, see  Section\eqref{quasihole}, correspond to the degenerate representations:
\begin{equation}
\sigma=\Phi_{(2,1,\dots,1|1,1,\dots,1)} \qquad \sigma' = \Phi_{(1,,1,\dots,2|1,1,\dots,1)}
\end{equation}
whose dimensions is given by \eqref{hsigma}. The $W^{(3)}_0$ eigenvalues have been fixed to:
\begin{equation}
\omega^{(3)}_{\sigma}=  -\omega^{(3)}_{\sigma'} = \sqrt{ \frac{(k-1)^2(k-2)}{18 k^3} \frac{(2k+1-kr)^2(3k+2-kr)}{(k+r)^2(r(k+2)+k)}}
\label{w3sigma}
\end{equation}
We would like to stress that the two sets of fields $(\Psi_1,\Psi_{k-1})$ and $(\sigma,\sigma')$ are dual from  one another under the transformation 
\begin{equation}
\textrm{WA}(p,q)\rightarrow \textrm{WA}(q,p) \label{CFTduality}
\end{equation}
i.e. $\alpha_{+}\leftrightarrow \alpha_{-}$. For the minimal model $\textrm{WA}_{1}(p,q)$ this duality is usually expressed in terms of an electric-magnetic duality transformation, also called T-duality in the literature \cite{Wiegman}. Under this transformation the value of $r$ changes into its dual $\tilde{r}$:
\begin{equation}
 (k+r)(k+\tilde{r}) = (k+1)^2
 \label{duality}
\end{equation}
In particular one can check that the quantum numbers $(\Delta,(w^{(3)}_{\Psi})^2)$ and  $(h,(w^{(3)}_{\sigma})^2)$ are exchanged under this transformation.

\section{Second order differential equations for ground state and quasihole wavefunctions }
\label{PDE}

In \cite{RB1} the $N$-points correlation function $\langle\Psi(z_1) \Psi(z_2)\cdots \Psi(z_N)\rangle^{(r)}$  of the $\textrm{WA}_{k-1}(k+1,k+r)$ theory has been considered. This correlation function gives the ground state of the corresponding $(k,r)$ Jack FQH state.
By using the Ward identities associated to the spin $3$ current $W^{(3)}(z)$ and the degeneracy properties of the $\Psi_{1}$ and $\Psi_{k-1}$ representations, it was  showed that their $N-$points  correlation functions  satisfies a second order differential equation.  In particular, this equation can be transformed into a Calogero-Sutherland Hamiltonian with negative rational coupling $\alpha=-(k+1)/(r-1)$. This completed the proof of the conjecture  \cite{  BernevigHaldane2,FJMM,FJMM2, Mathieu,BernevigW}  which states that the $N-$points correlation functions of $\Psi_{1}$ ($\Psi_{k-1}$) can be written in term of a single Jack polynomial. 

In this section we  extend this analysis to general quasihole wavefunctions, and we derive a partial differential equation satisfied by the correlator $\langle \sigma(w_1) \sigma(w_2)\cdots \sigma(w_M) \Psi(z_1)\Psi(z_2)\cdots \Psi(z_N)\rangle$.

\subsection{General relations  from Ward identities}

 The $\textrm{WA}_{k-1}$ symmetry manifests itself at the quantum level in the Ward identities associated to  the symmetry currents $T(z)$ and $W^{(s)}(z)$, $s=3,\cdots,k$.  
 In \cite{RB1},  we have written down these relations for the specific case of the $N-$points $\Psi$ correlation function. These relations are very general and apply to  any primary fields  correlation function of the $\textrm{WA}_{k-1}$ theory.

 These identities  can be easily obtained from \eqref{Wprimary}.   Consider a set of $W$ primaries $\phi_i$ of dimension $\Delta_{\phi_i} $ and $W^{(3)}_0$ eigenvalues $\omega^{(3)}_{\phi_i}$. The Ward identities associated to the 
   the stress energy tensor $T(z)$ and $W^{(3)}(z)$ take the form:
\begin{eqnarray}
\langle T(z) \phi_1(z_1)\cdots \phi_N (z_N)\rangle &=& \sum_{j=1}^{N} \left(\frac{\Delta_{\phi_j}}{(z-z_j)^2} + \frac{\partial_j}{(z-z_j)} \right)\langle \phi_1(z_1) \cdots \phi_N(z_N) \rangle   \label{Ward_vir}\\
\langle W^{(3)}(z) \phi_1(z_1)\cdots \phi_N(z_N) \rangle &=& \sum_{j=1}^{N} \left( \frac{\omega^{(3)}_{\phi_j}}{(z-z_j)^3} \langle \phi_1(z_1)\cdots \phi_N(z_N)\rangle +  \langle \phi_1(z_1)\cdots \frac{W_{-1}^{(3)}\phi_j(z_j)}{(z-z_j)^2}\cdots \phi_N(z_N)\rangle  \right.\nonumber \\
&&+ \left.\langle \phi_1(z_1)\cdots  \frac{W^{(3)}_{-2}\phi_j(z_j)}{(z-z_j)} \cdots \phi_N (z_N)\rangle \right)
\label{Ward_W}
\end{eqnarray}
The asymptotics of the functions $\langle T(z)\dots\rangle$ and $\langle W^{(3)}(z)\dots\rangle$ have  respectively the form:
\begin{equation}
T(z)\sim \frac{1}{z^4} \quad \textrm{ and } \quad W^{(3)}(z) \sim \frac{1}{z^6} \quad \textrm{as } z\to \infty
\label{asymp}
\end{equation}
These asymptotic behaviors, together with the Ward identities \eqref{Ward_vir}-\eqref{Ward_W}, imply a set of relations satisfied by the correlation functions $\langle\phi_{1}\cdots \phi_N\rangle$. For instance, plugging the decomposition \eqref{Ward_vir} into \eqref{asymp}, one finds  simple differential equations which impose the invariance of the correlation function $ \langle \phi_1(z_1)\cdots \phi_N(z_N) \rangle$ under global conformal transformations \cite{diFrancesco}.  Analogously to the case of the conformal symmetry, one can derive a set of relations associated to the symmetry generated by the spin $3$ current $W^{(3)}(z)$  \cite{RB1,TFTcf}. In particular we will use the following relation:
\begin{equation}
\sum_{j=1}^{N}  \langle \phi_1(z_1) \cdots \left(z^2_j W_{-2}^{(3)} +2 z_j W_{-1}^{(3)}+\omega^{(3)}_{\phi_j} \right)\phi_j(z_j)\cdots \phi_N(z_N) \rangle =0 \label{w3}\\
\end{equation}

\subsection{Second level null vector conditions.}
In \cite{RB1} it was showed that, for the first two levels, all the $W$ descendants of  the fields $\Psi_{1}$ and $\Psi_{k-1}$ can be expressed in terms of Virasoro modes. As already mentioned, the fields $\sigma$ and $\sigma'$ are dual to the fields $\Psi_1$ and $\Psi_{k-1}$ in the sense of \eqref{duality}. This means also that, apart from different values of their quantum numbers, the associated representation modules have the same structure, and in particular the same degeneracies. These four primary fields have null vectors of the form:
\begin{eqnarray}
W^{(3)}_{-1} \Phi_i & = &a_i L_{-1} \Phi_i  \nonumber \\
W^{(3)}_{-2} \Phi_i & = & \left(\mu_i L_{-1}^2 +\nu_i L_{-2} \right)\Phi_i 
\label{null_vector_generic}
\end{eqnarray}
where the exact coefficients $a_i,\mu_i$ and $\nu_i$ depend on the quantum numbers of the field under consideration:
\begin{eqnarray}
a_i & = & \frac{3\omega^{(3)}_i}{2\Delta_i} \nonumber \\
\mu_i & = & a_i\frac{2(2\Delta_i+c)}{(-10\Delta_i+16\Delta_i^2+2c\Delta_i+c)} \nonumber   \\
\nu_i & = & \mu_i \frac{8\Delta_i(\Delta_i-1)}{(2\Delta_i+c)} 
\label{coeff_null}
\end{eqnarray}

\subsection{From the Ward identity to differential equations for conformal blocks}

Following \cite{RB1,TFTcf,TFTcf2}, we derive a second-order differential equation satisfied by all the conformal blocks $\langle \sigma(w_1)\cdots \sigma(w_M) \Psi(z_1) \cdots \Psi(z_N)\rangle$.  In order  to write any of the relations \eqref{w3} in a differential form, we use the null-vector conditions \eqref{null_vector_generic} to express the action of the modes $W^{(3)}_{-2}$ and $W^{(3)}_{-1}$ in terms of the  Virasoro modes $L_{-2}$ and $L_{-1}$($=\partial$).  Notice that  one could start from another  Ward identity instead of \eqref{w3}, and obtain a different differential equation.  As suggested by the results known for the Jacks \cite{Bernevig}, all these differential equations are not independent and can be obtained form one another.

We consider the most general conformal block describing a $M$ quasiholes wavefunction \eqref{conformal block label}:
\begin{equation}
\mathcal{F}^{(r)}_{(a)}(\{w_i\},\{z_i\})=\langle \sigma(w_1)\cdots\sigma(w_M)\Psi_1(z_1)\cdots\Psi_1(z_N)\rangle_{(a)}^{(r)} \label{qhcorr}
\end{equation}
By using \eqref{null_vector_generic} in \eqref{w3}, we obtain the following PDE for  $\mathcal{F}^{(r)}_{(a)}$:
\begin{equation}
\mathcal{H}(k,r)\mathcal{F}^{(r)}_{(a)}(\{w_i\},\{z_i\})=0
\label{de}
\end{equation}
where the differential operator $\mathcal{H}(k,r)$ is a second order differential operator depending on the two integers $k,r$ parametrizing the theory $\textrm{WA}_{k-1}(k+1,k+r)$. 
Defining:
\begin{equation}
L_{-2}^{(x_i)}\langle \phi_1(x_1)\cdots\phi_N(x_N)\rangle=\left(\sum_{j\neq i}\frac{\Delta_j}{(x_i-x_j)^2}+\frac{\partial_{x_j}}{(x_i-x_j)}\right) \langle \phi_1(x_1)\cdots\phi_N(x_N)\rangle,
\end{equation}
the differential operator $\mathcal{H}(k,r)$ has the following form:
\begin{eqnarray}
\mathcal{H}(k,r)& = &  \mu_{\Psi}\sum_{i=1}^N \left[ z_i^2 \left( \partial_i^2 + \frac{\nu_{\Psi}}{\mu_{\Psi}} L_{-2}^{(z_i)} \right)+ \frac{a_{\Psi}}{\mu_{\Psi}} \left( 2z_i \partial_{z_i} + \frac{2}{3}\Delta \right) \right] \nonumber \\
& + &\mu_{\sigma}  \sum_{l=1}^M \left[ w_l^2 \left( \partial_{w_l}^2 + \frac{\nu_{\sigma}}{\mu_{\sigma}} L_{-2}^{(w_l)} \right)+ \frac{a_{\sigma}}{\mu_{\sigma}} \left( 2w_l \partial_{w_l} + \frac{2}{3}h \right) \right] 
\label{hform}
\end{eqnarray}
The coefficients $\mu_{\Psi,\sigma}$ and $\nu_{\Psi,\sigma}$, which depends on $k$ and $r$, are defined in \eqref{coeff_null}. 

If we perform the addition of an arbitrary number of unit of flux on a Jack FQH ground-state, we obtain in general a degenerate space spanned by all the quasihole conformal blocks. In the CFT context, it is clear that all these different conformal blocks obey the same PDE, and it is manifestly the case in \eqref{hform}. As such this PDE is quite complicated and not very practical. In particular the operators $L_{-2}^{(z_i)}$ and $L_{-2}^{(\omega_l)}$ mixes electron and quasihole coordinates. 

However this PDE acts on CFT conformal blocks and not the actual FQHE wavefunctions, which differ by a $\textrm{U}(1)$ charge term. In the next section we show that taking into account this additional charge term greatly simplifies this differential equation.

\section{PDE for quasihole wavefunctions and  quasihole - electron duality}

In this section the main results of this paper are presented. We derive the PDE for generic quasihole wavefunctions describing the non-abelian Jack quantum Hall states. This PDE turns out to be surprisingly simple, and exhibits an interesting duality between electrons and quasiholes. Although electrons and quasiholes have very different physcial properties, the differential equation is invariant under the exchange:
\begin{equation*}
\Psi \leftrightarrow \sigma, \qquad r  \leftrightarrow \tilde{r}= \frac{2k+1-kr}{k+r}
\end{equation*}
We then use this PDE to compute the most generic wavefunction with one extra flux quantum. We recall that, by setting $r=2$ in the above equations, one finds the second order differential equations satisfied by the $k$ Read-Rezayi quasihole wavefunctions. 

\subsection{PDE and Calogero-Sutherland Hamiltonians}

We consider the $(k,r)$ Jack wavefunction for $M$ quasiholes and $N$ electrons, as defined through the conformal block:
\begin{equation}
F(\{\omega\},\{z\}) \equiv \langle \sigma(\omega_1) \cdots \sigma(\omega_M) \Psi(z_1)\cdots \Psi(z_N)\rangle \prod_{i<j} (z_i-z_j)^{\frac{r}{k}}\prod_{l<m} (\omega_l-\omega_m)^{\frac{\tilde{r}}{k}} \prod_{i,l}(z_i-\omega_l)^{\frac{1}{k}} \label{qhwf}
\end{equation}
Note that this is not exactly the usual wavefunction, as the quasihole $\textrm{U}(1)$ term has been taken to be $\prod_{k<l}(\omega_l-\omega_m)^{\frac{\tilde{r}}{k}}$ rather than $\prod_{k<l}(\omega_l-\omega_m)^{\frac{1}{2rk}}$. This choice of normalization was made in order to emphasize the quasihole-electron duality. Taking into account the additional $\textrm{U}(1)$ term in \eqref{qhwf}, the PDE \eqref{hform} simplifies into:
\begin{equation}
\alpha \mathfrak{H} ^{(\alpha)}(z) F(\{\omega\},\{z\}) =  \tilde{\alpha} \mathfrak{H} ^{(\tilde{\alpha})}(\omega) F(\{\omega\},\{z\})  \label{double_CS}
\end{equation}
where $\mathfrak{H} ^{(\alpha)}(z)$ and $\mathfrak{H} ^{(\tilde{\alpha})}(\omega)$ are differential operators of order two acting respectively on electron and quasihole coordinates. Moreover they both belong to the Calogero-Sutherland family of commuting Hamiltonians, but for different coupling $\alpha$ and $\tilde{\alpha}$:
\begin{eqnarray}
\mathfrak{H} ^{(\alpha)}(z) &\equiv & \left( \mathcal{H}^{(\alpha)}_{\textrm{CS}}(z) - \epsilon^{(0)}(r,N) \right) + \left(\frac{N-k}{\alpha} -1 \right)\left( \sum_{i=1}^N z_i \partial_{z_i} - \frac{1}{2}N N_{\Phi}^{(0)}(r,N) \right) - \frac{MN}{k^2}(M-k) \label{CS_electron}\\
\mathfrak{H} ^{(\tilde{\alpha})}(\omega) &\equiv & \left( \mathcal{H}^{(\tilde{\alpha})}_{\textrm{CS}}(\omega) -  \epsilon^{(0)}(\tilde{r},M) \right) + \left(\frac{M-k}{\tilde{\alpha}} -1 \right)\left( \sum_{l=1}^M \omega_l \partial_{\omega_l} - \frac{1}{2}M N_{\Phi}^{(0)}(\tilde{r},M) \right) - \frac{MN}{k^2}(N-k) \label{CS_qh}
\end{eqnarray}
where we introduced the electronic Calogero-Sutherland coupling $\alpha$ and the dual quasihole coupling $\tilde{\alpha} = 1- \alpha$, i.e.
\begin{equation}
\alpha = -\frac{k+1}{r-1} \qquad \tilde{\alpha} = \frac{k+r}{r-1} 
\end{equation}
and the following quantities:
\begin{itemize}
\item the ground state number of flux $N_{\phi}^{(0)}(r,N) \equiv r \frac{N-k}{k}$
\item the Calogero-Sutherland eigenvalue  $\epsilon_{\lambda}^{(0)}(r,N)  \equiv  \frac{N r (N - k) [2 N r + k^2 (1 - 2 r) + k (N - r + N r)]}{6 k^2 (k+1)}$ corresponding to the ground state partition $\lambda^{(0)} = [k0^{r-1}k0^{r-1}k\cdots 0^{r-1}k]$ and the coupling constant $\alpha = -\frac{k+1}{r-1}$
\end{itemize}
The form \eqref{double_CS} of the PDE is extremely simple, as it does not mix electron and quasihole coordinates. Moreover it extends the proof relating $\textrm{WA}_{k-1}(k+1,k+r)$ theories to $(k,r)$ admissible Jack polynomials to the generic case of quasihole wavefunctions. 

Even more interestingly, this PDE manifestly invariant under the transformation: 
\begin{equation*}
\Psi \leftrightarrow \sigma, \qquad \alpha \leftrightarrow 1-\alpha \qquad \textrm{i.e.} \qquad r   \leftrightarrow \tilde{r}= \frac{2k+1-kr}{k+r}
\end{equation*}
This is induced by the CFT duality \eqref{duality}, which translates in this context into an electron-quasihole duality at the PDE level. Notice however that this is not a symmetry of the quantum Hall state, since electrons and quasiholes have very different physical properties.

\section{Full expansion of the $k$-quasihole wavefunction}

We have directly verified that all  the known RR wavefunctions proposed in earlier works satisfy \eqref{double_CS}.  These differential equations provide a rigorous proof of the validity of the ansatz used in \cite{Ardonne_Schoutens} to compute the four quasihole wavefunctions for the general $k$ Read-Rezayi states.

As an explicit application of this differential equation, we compute in the following the full expansion of the $1$-flux added $k$ quasihole wavefunctions for any Jack state, including the Read-Rezayi sequence. The full expansion of these wavefunctions were previously unknown, and illustrates the electron-quasihole duality. It also provides a proof of the form of the $k$-quasihole Jack wavefunction with $k-1$ quasiholes at  the north pole proposed in \cite{BernevigHaldane2}, which was obtained assuming that the electronic part was spanned by $(k,r)$ admissible Jack polynomials.  

In appendix \ref{app_expansions}  we provide the expansions of all other $k$-quasihole wavefunctions with $n$ quasiholes at the north pole and $(k-n)-1$ quasiholes at the south pole, as obtained from the clustering condition. All these expressions are recovered by an appropriate specialization of the full expansion derived in this section.

\subsection{Structure of the $k$-quasihole wavefunction}

In the presence of a single extra flux, $k$ elementary quasiholes  are present in the quantum Hall liquid.  For $M=k$ quasiholes and $N$ electrons, the CFT correlator \eqref{qhcorr} is single channeled and the wavefunction $F(\{w\},\{z\})$ as defined in \eqref{qhwf} is simply a polynomial in $\{w,z\}$. The quasihole part of the wavefunction can be expanded into the eigenbasis of $\mathcal{H}^{(\tilde{\alpha})}_{CS}$, namely Jack polynomials $J^{(\tilde{\alpha})}_{\mu}(\{w\})$
\begin{equation}
F(\{w\},\{z\}) = \sum_{\mu} a_{\mu} J_{\mu}^{\tilde{\alpha}}(\{w\})  P_{\mu}(\{z\})  \label{decomp_qh}
\end{equation} 
The sum over $\mu$ means the sum over all $\times k$ partitions $\frac{N}{k}  \geq \mu_1\geq \mu_2 \cdots \mu_k \geq 0$. The corresponding Jacks $J_{\mu}^{\tilde{\alpha}}(\{w\})$ span the entire quasihole space ($k$ bosons in $\frac{N}{k}$ orbitrals), and are always well defined  as $\tilde{\alpha} = \frac{k+r}{r-1}$ is positive.
Plugging this expression in the PDE \eqref{double_CS}, we obtain that $P_{\mu}(\{z\})$ is an eigenvector of  $\mathcal{H}^{(\alpha)}_{CS}$. Neglecting the accidental spectrum degeneracies  of $\mathcal{H}^{(\alpha)}_{CS}$, $P_{\mu}(\{z\})$  is then a Jack polynomial $J_{\lambda(\mu)}^{(\alpha)}$ for some partition $\lambda(\mu)$ depending on $\mu$:
\begin{equation}
F(\{w\},\{z\}) = \sum_{\mu} a_{\mu} J_{\mu}^{\tilde{\alpha}}(w)  J_{\lambda(\mu)}^{\alpha}(z)  \label{decompJack}
\end{equation} 
To any partition $\mu$ corresponds a $(k,r)$ admissible partition $\lambda(\mu)$ . The correspondence $\mu \to \lambda$ is determined by plugging the function  $J_{\lambda}^{\alpha}(z)  J_{\mu}^{\tilde{\alpha}}(w)$ in the PDE \eqref{double_CS}, which leads to the following equations between eigenvalues:
\begin{equation}
\epsilon_{\lambda}(\alpha) - \epsilon_{\lambda}^{(0)}(\alpha) + \frac{k+r}{k+1}\left( \epsilon_{\mu}(\tilde{\alpha}) -\frac{N^2}{k} \right) + \frac{r-1}{k+1}(N-k-1) \left(  \sum_{i=1}^k \mu_i - N \right) = 0 \label{eigenvalues_relation}
\end{equation}
The inverse mapping $\lambda \to \mu$ is the following: to the $(k,r)$ admissible partition $\lambda$ corresponds the quasihole partition $\mu$ defined as :
\begin{equation}
 \mu_i = \sum_{\substack{ n=1 \\n =k+1 -i \textrm{ mod } k}}^N \left( 1 - l_n \right) \qquad \textrm{for } i=1,\cdots, k \label{mapping}
\end{equation}
where we introduced for compactness $l_i \equiv \lambda_i - \lambda_i^{(0)}$ for $i=1,\cdots, N$ ($\lambda^{(0)}$ is the densest $(k,r)$ admissible partition). Note that $l_i \in \{0,1\}$ when there is a single extra flux, and this is why this case is quite simple (there is no trace of the non abelian braiding yet, and the counting of $(k,r)$ admissible partitions is trivial). \\
Equivalently the direct mapping is:
\begin{equation}
(l_i,  l_{i+k}, l_{i+2k},\cdots, l_{i+ N-k}) = ( \underbrace{1,1,1,\cdots 1}_{N/k-\mu_{k+1-i}}, \underbrace{0,\cdots,0}_{\mu_{k+1-i}}) \label{direct_mapping}
\end{equation} 
It is straightforward to check that this mapping ensures the relation \eqref{eigenvalues_relation}.

\subsection{Global translation invariance}

The explicit expression of the coefficients $a_{\mu}$'s in \eqref{decompJack} can be obtained by demanding global translation invariance of the quasihole wavefunction and using Lassale's formula for Jack polynomials.
\begin{equation}
\left(\sum_{i=1}^N \partial_{z_i} + \sum_{l=1}^k \partial_{\omega_l} \right) \left(\sum_{\mu} a_{\mu} J_{\mu}^{(\tilde{\alpha})}(w) J_{\lambda}^{(\alpha)}(z) \right)= 0 
\end{equation}
When acting with  $L_- = \sum_i \partial_{z_i} $ on a Jack polynomial $J_{\lambda}^{(\alpha)}$ we get :
\begin{equation}
L_- J_{\lambda}^{(\alpha)} = \sum_i A(\lambda,\lambda(i),\alpha) J_{\lambda(i)}^{(\alpha)}
\end{equation}
where the partition $\lambda(i)$ is obtained from $\lambda$ by removing a box in the  i$^{th}$ row, and the coefficient $A(\lambda,\lambda(i),\alpha)$ is given by Lassale's formula \cite{Lassale}:
\begin{equation}
A(\lambda,\lambda(i),\alpha) = \frac{N-i+\lambda_i\alpha}{\alpha} \prod_{j=i+1}^N \frac{(\alpha(\lambda_i-\lambda_j-1)+j-i+1)(\alpha(\lambda_i-\lambda_j)+j-i-1)}{(\alpha(\lambda_i-\lambda_j)+j-i)(\alpha(\lambda_i-\lambda_j-1)+j-i)}
\end{equation}
When acting with  $L_-(\{w\}) + L_- (\{z\}) =\sum_l \partial_{w_l} + \sum_i \partial_{z_i} $ on 
\begin{equation}
F(\{w\},\{z\})= \sum_{\mu} a_{\mu} J_{\mu}^{\tilde{\alpha}}(\{w\})  J_{\lambda(\mu)}^{\alpha}(\{z\})
 \end{equation} 
we obtain a polynomial of the form:
\begin{equation}
\left(\sum_{i=1}^N \partial_{z_i} + \sum_{l=1}^k \partial_{\omega_l} \right)  F(\{w\},\{z\}) = \sum_{\mu} \sum_i \left(  a_{\mu(i)} A(\mu,\mu(i),\tilde{\alpha}) + a_{\mu} A(\lambda^{(i)},\lambda,\alpha)\right)J_{\mu(i)}^{\tilde{\alpha}}(w)  J_{\lambda(\mu)}^{\alpha}(z) 
\end{equation} 
where $\mu(i)$ is the partition obtained from $\mu$ and removing a box in the i$^{th}$ row, 
$\lambda = \lambda(\mu)$ and $\lambda^{(i)} \equiv \lambda(\mu(i))$ is the partition corresponding to $\mu(i)$ through the mapping \eqref{mapping}. Explicitly the partition $\lambda^{(i)}$ is obtained from $\lambda$ by adding a box in the $(N+k+1-i-k\mu_i)^{\mbox{th}}$ row:
\begin{equation}
\lambda^{(i)}_j = \lambda_j + \delta_{j,N+(k+1-i)-k\mu_i} 
\end{equation}
Translation invariance boils down to the set of relations:
\begin{equation}
a_{\mu(i)} = -  a_{\mu} \frac{ A(\lambda^{(i)},\lambda,\alpha)}{A(\mu,\mu(i),\tilde{\alpha}) } 
\end{equation}
The generic expressions for $A(\mu,\mu(i),\tilde{\alpha})$ and $A(\lambda^{(i)},\lambda,\alpha)$ are quite complicated. However  their ratio turns out to be extremely simple:
\begin{equation}
\frac{A(\mu,\mu(i),\tilde{\alpha})}{A(\lambda^{(i)},\lambda,\alpha) } = \frac{k+1-i}{i}
\end{equation}
Up to a global multiplicative constant, we find 
\begin{equation}
a_{\mu} = \prod_{i=1}^{k} \left( - \frac{k+1-i}{i}\right)^{\mu_i} \label{general_coeff}
\end{equation}
This is quite remarkably independent of $r$: these coefficients are the same for  the RR states and their $r>2$ generalizations. To summarize we obained the most general $k$ quasiholes wavefunction for the $(k,r)$ Jack states to be:
\begin{equation}
F(\{w\},\{z\})= \sum_{\mu} a_{\mu} J_{\mu}^{\tilde{\alpha}}(\{w\})  J_{\lambda(\mu)}^{\alpha}(\{z\}) \qquad \qquad a_{\mu} = \prod_{i=1}^{k} \left( - \frac{k+1-i}{i}\right)^{\mu_i} \label{kqh_wv}
 \end{equation} 
with $\alpha = 1-\tilde{\alpha} = -\frac{k+1}{r-1}$. 

This expression generalizes the expansions obtained in  \cite{BernevigHaldane2}, which correspond to the specific case of  $k-1$ quasiholes located at the north pole (see Appendix \ref{app_expansions}).

\section{Conclusion}
In this paper we have derived a second-order PDE, written in \eqref{hform},  satisfied by the most general Read-Rezayi and Jack  quasihole wavefunctions. In order to achieve this result,  we have generalized the results of \cite{RB1}, in which the ground state wavefunctions were considered, to the quasihole excited wavefunctions.  We showed that this is possible because  the $\textrm{WA}_{k-1}$ representation modules corresponding to the quasihole and the particle operators share the same degeneracy structure. The $W$ symmetry is completely manifest in  these differential equations, thus completing the observations done in \cite{Cappelli},  which  explicitly depend on the values of the $W^{(3)}_0$ eigenvalues.

The differential equation decomposes as the sum of two Hamiltonians of Calogero-Sutherland type  acting on the electron and quasihole coordinates, unravelling  a remarkable electron-quasihole duality at the PDE level.
We used these results to characterize the quasihole wavefunctions whose explicit expression is in general not known for these theories.  We proved the validity of some ansatz used in \cite{Ardonne_Schoutens} and in  \cite{BernevigW} to compute certain quasihole wavefunctions. Finally we provided  the full expansion  of the $1$-flux added  quasihole wavefunctions.   
On the basis of this PDE, it would be interesting to understand the quasihole wavefunctions with more than one flux added, and their non-Abelian monodromy properties, in terms of non-polynomial solutions of Calogero-Sutherland Hamiltonians. This is, at our knowledge, a quite unexplored mathematical problem.

As a last remark, we point out that  the question of the vanishing of the Berry connection of the  Read-Rezayi states has not been completely proven, even if  general conditions for it to hold has been discussed in \cite{Read_holonomy}. In this respect,  some strong arguments for the Moore-Read states have been provided by using the Coulomb gas approach \cite{Gurarie_Nayak}.  For  Halperin's Abelian spin-singlet state, it was shown in \cite{Block_Wen} 
that the Berry connection vanishes  by using the Knizhnik-Zamolodchikov equations. It should be interesting then to explore the possibilities to compute the Berry connection for the Read-Rezayi states by using the $\textrm{WA}_{k-1}$ Coulomb gas representation or by using the differential equations we derived in this paper.

{\it Acknowledgements}:

We have greatly benefited from discussions with E. Ardonne and N.Regnault.
 B.Estienne was supported by FOM of the Netherlands. B.A.B was supported by Princeton University Startup Funds and by the Alfred P. Sloan Foundation. B.A.B wishes to thank the Institute of Physics Center for International Collaboration in Beijing, China for generous hosting.

\section{Appendix: expansions of some specialized $k$-quasihole wavefunctions}
\label{app_expansions}

The expansion of the $k$-quasihole wavefunction with $k-1$ quasiholes bunched up at the north pole in terms of Jack polynomials has been proposed in \cite{BernevigHaldane2}.  The derivation was based on the specific clustering properties of this wavefunction, and relied on the assumption that only $(k,r)$ Jack polynomials appear in the expansion.  Using a similar approach, we derive in this section the expansion of the wavefunctions with $n$ quasiholes at the north pole and $(k-n)-1$ quasiparticles at the south pole in terms of powers of the remaining quasihole. On the plane, this corresponds to the wavefunction with $n$ quasiholes at zero and $(k-n)-1$ at infinity. We then compare the expression found by the clustering method with the special cases of the general expressions obtained in the algebraic calculations in this paper.

The physical operation we can perform on a FQH ground-state is the addition of one unit of flux, which corresponds, in this case, to the addition of $k$ charge $\frac{1}{k}$ quasiholes, which then further fractionalize and become distinct.  We then take $n$ quasiholes to the north pole (origin in the disk geometry), $(k-n)-1$ at the south pole (infinity in the disk geometry)  and the $k$'th
non-Abelian quasihole at $w$. This fractionalized quasihole wavefunction $\psi(w; z_1..z_N)$ is defined by the following clustering property:
\begin{equation}
\psi_n(w; z_1,\cdots, z_N)|_{z_1 =\cdots=z_k =w} =0
\end{equation}
\noindent which destroys the wavefunction if $k$ particles reach the fractionalized quasihole $w$. We impose two further conditions:
\begin{itemize}
\item  $\psi_n(w; z_1,\cdots, z_N)|_{z_1 =\cdots=z_{k-n+1}=\text{North Pole} } =0$ pins $n$ quasiholes at the North pole,
\item $\psi_n(w; z_1,\cdots, z_N)|_{z_{1} =\cdots=z_{n+2}=\text{South Pole}} =0$ pins $(k-n)-1$ quasiparticles at the South pole
\end{itemize}
When zero quasiholes are either at the North or South pole, the two conditions become identical to the electron clustering properties of a $(k,r)$ FQH state. We can most easily express this wavefunction as a function of the difference between the number of quasiholes at the north and south pole: $\Delta n= n -((k-n)-1) = 2 n - k+ 1$, which we take without any loss of generality to be $\ge 0$. The $\le0$ part of $\Delta n$ is just a mirror reflection of the $\Delta n\ge0$. We have
\begin{equation}
\psi_n(w  ; z_1,\cdots z_N )= \sum_{i=0}^{N/k} (- \beta_n  w )^i  J^{\alpha}_{\lambda_i^{(n)}}(\{z\})
\end{equation}
where 
\begin{equation} \beta_n = \frac{k+1 - \Delta n}{k+1+ \Delta n} = \frac{k-n}{n+1}\end{equation} and the polynomials $J^{\alpha}_{\lambda_i^{(n)}}(\{z\})$ are Jacks  with partition $\lambda_i^{(n)}$ given by, in orbital occupation notation:  
\begin{eqnarray}
&\lambda_0^{(n)}= |a_n \,b_n \,0\cdots0 \, a_n\,b_n \,0\cdots0 \,a_n\,b_n\, 0 \cdots0\,a_n\,b_n  \rangle,\quad \nonumber \\
& \lambda_1^{(n)}= |a_n+1\,b_n-1\, 0  \cdots 0\, a_n\,b_n \,0 \cdots0\, a_n\,b_n \, 0\cdots 0\,a_n\,b_n \rangle, \quad\nonumber \\
 & \lambda_2^{(n)} = |a_n+1\,b_n-1\, 0\cdots0 \,a_n+1\,b_n-1\, 0\cdots0 \, a_n\,b_n \, 0\cdots0 \,a_n\,b_n \rangle, \quad
\nonumber \\
 &\lambda_3^{(n)} = |a_n+1\,b_n-1\, 0\cdots 0 \,a_n+1\,b_n-1 \,0\cdots 0 \,a_n+1\, b_n-1\,0 \cdots 0\,a_n\,b_n  \rangle, \quad \nonumber \\
  & \cdots \nonumber \\ 
  & \lambda_{\frac{N}{k}}^{(n)} = |a_n+1\,b_n-1 \,0 \cdots 0 \,a_n+1\,b_n-1 \,0 \cdots 0\, a_n+1\, b_n-1\,0 \cdots 0\,a_n+1\,b_n-1\rangle \label{clustered_partitions}
\end{eqnarray} where
\begin{equation}
a_n= \frac{k-1 - \Delta n}{2} = k-n-1\qquad b_n = n+1
\end{equation}
and where the number of zeroes between one $b_n$ and the next $a_n$ to the right of it is $r-2$. Of course, the kets above are admissible Jacks at $-(k+1)/(r-1)$; the partitions are obviously $(k,r)$-admissible. 

The  wavefunctions obtained above are all special cases of  the generic $k-$quasiholes wavefunction \eqref{kqh_wv}, and they can be recoverd by putting $n$ quasihole at $0$ and $k-n-1$ at infinity:
\begin{equation}
\psi_n(w ; z_1,\cdots,z_N) = F(w_1 = \cdots = w_n = 0, w_{n+1} = w, w_{n+2} = \cdots = w_k = \infty ; z_1, \cdots z_N )
\end{equation}
This specialization kills all the $J_{\mu}^{\tilde{\alpha}}(\{w \})$, except for the partitions
\begin{eqnarray}
& \mu^{(n)}_0 = \big[ \overbrace{\textstyle\frac{N}{k}\cdots \frac{N}{k}}^{k-n-1}\,0\,\overbrace{\textstyle0\vphantom{\frac{N}{k}}\cdots 0}^{n}\big] \nonumber \\
& \mu^{(n)}_1 = \big[ \frac{N}{k}\cdots \frac{N}{k}\,1\,0\cdots 0 \big] \nonumber \\
& \mu^{(n)}_2 = \big[ \frac{N}{k}\cdots \frac{N}{k}\,2\,0\cdots 0 \big] \nonumber \\
& \cdots \nonumber \\
& \mu^{(n)}_{\frac{N}{k}} = \big[ \frac{N}{k}\cdots \frac{N}{k}\,\frac{N}{k}\,0\cdots 0 \big] \nonumber \\
\end{eqnarray} 
for which the quasihole Jack polynomial becomes
\begin{equation}
J_{\mu^{(n)}_i}^{\tilde{\alpha}} \left( w_1=\cdots= w_{n} = 0 ; w_{n+1}=w, w_{n+2} =\cdots= w_{k} =\infty \right) = w^{i}
\end{equation}
According to the mapping \eqref{direct_mapping}, the corresponding electronic partitions $ \lambda(\mu_i^{(n)})$ are precisely the partitions \eqref{clustered_partitions} :
\begin{equation}
 \lambda(\mu_i^{(n)}) = \lambda_i^{(n)}  \qquad i=0,\cdots \frac{N}{k}
\end{equation}
and we aslo recover the value of $\beta_n$ from the general value of $a_{\mu}$ in \eqref{general_coeff}:
\begin{equation}
a_{\mu^{(n)}_i} \propto \left(-\frac{k-n}{n+1}\right)^i \qquad \Rightarrow \qquad \beta_n =  \frac{k-n}{n+1}
\end{equation}

\end{document}